\documentclass{article}
\usepackage{spconf,amsmath,graphicx,chngpage}
\usepackage[utf8x]{inputenc}
\graphicspath{{images/}}

\usepackage{enumitem}
\setlist{nosep, leftmargin=14pt}
\usepackage{mwe} 

\title{Diagnosis of sickle cell anemia using AutoML on UV-Vis absorbance spectroscopy data}
\name{Sarthak Srivastava$^{1}$, Radhika N. K.$^{1}$, Rajesh Srinivasan$^{1}$, Nishanth K M Nambison$^{2}$, and Sai Siva Gorthi$^{1*}$}
\address{Indian Institute of Science Bangalore$^{1}$\\ Government Homeopathic Medical College \& Hospital, Bhopal$^{2}$\\ *Corresponding author email: saisiva@iisc.ac.in}

\begin{document}
%
\maketitle
\begin{abstract}
Sickle cell anemia is a genetic disorder that is widespread in many regions of the world. 
Early diagnosis through screening and preventive treatments are known to reduce mortality in the case of sickle cell disease (SCD). In addition, the screening of individuals with the largely asymptomatic condition of sickle cell trait (SCT) is necessary to curtail the genetic propagation of the disease. However, the cost and complexity of conventional diagnostic methods limit the feasibility of early diagnosis of SCD and SCT in resource-limited areas worldwide. Recently, our group developed a low-cost UV-Vis absorbance spectroscopy based diagnostic test for SCD and SCT. Here, we propose an AutoML based approach to classify the raw spectra data obtained from the developed UV-Vis spectroscopy technique with high accuracy. The proposed approach can detect the presence of sickle hemoglobin with 100\% sensitivity and 93.84\% specificity. This study demonstrates the potential utility of the machine learning-based absorbance spectroscopy test for deployment in mass screening programs in resource-limited settings.

\end{abstract}

\section{Introduction}
\label{sec:intro}

Hemoglobin is an oxygen-binding protein found inside red blood cells (RBCs). It is made up of four peptide chains, each bound to a heme group.
Hemoglobin A (or adult hemoglobin), which makes up 95--98\% of all hemoglobin content in normal healthy adults, is composed of two $\alpha$-globin and two $\beta$-globin peptide chains. In sickle cell disease, mutations in the HBB (hemoglobin subunit beta) gene which encodes for the $\beta$-globin peptide chains, gives rise to an abnormal form of hemoglobin called sickle hemoglobin (HbS) \cite{kato2018sickle}. Under oxidative stress, the HbS molecule polymerizes and forces the RBCs to assume a characteristic sickle shape. Sickle cell disease is characterized by debilitating pain, hemolysis, vaso-occlusion, stroke, and organ damage. Around 300,000 to 400,000 children globally are born each year with the condition. The disease is prevalent in sub-Saharan Africa and parts of the Mediterranean, the Middle East, and India.

Sickle cell anemia is an autosomal recessive disease, which means that an individual inherits sickle cell disease (SCD) only when both parents are carriers of the HBB gene mutation. In contrast, the carriers only possess a single copy of the mutated gene, and this condition is called sickle cell trait. Sickle cell trait is hard to detect as the carriers do not show any apparent health problems unless they are subjected to oxidative stress due to extreme conditions like dehydration or high altitude. Consanguineous marriages are still prevalent in southern Asia and sub-Sahara Africa leading to a large number of people being born with SCD. Mass screening programs to identify individuals with SCT and subsequent genetic counselling is the only way to reduce the prevalence of SCD. However, the standard diagnostic methods like HPLC (high-performance liquid chromatography) or IEF (isoelectric focusing electrophoresis) are expensive and technically complex. The dependence of these conventional tests on specialized equipment, stable 
infrastructure, and well-trained laboratory personnel are the major barriers limiting the deployment of large scale screening tests for SCT/SCD. In addition, the conventional solubility test, which is used for detection of sickle cell anemia, suffers from a high false positivity rate \cite{tubman2015sickle}.

In recent years, various low-cost methods have been proposed to detect SCD at the point-of-care \cite{piety2017paper, yang2013simple, nnodu2019hemotypesc}. Recently, Rajesh et al. developed a rapid low-cost optical-absorbance based test for the diagnosis of sickle cell disease and sickle cell trait \cite{rajesh}. The test is based on the differences in the UV-Vis absorbance spectra of normal hemoglobin and sickle hemoglobin suspended in a special buffer solution. The test criteria is based on the ratio of the absorbance at two peak wavelengths. However, this test criteria is sensitive to the instrument to instrument variations among spectrometers made by different manufacturers, necessitating the characterization and calibration of the spectrometers prior to testing. 

To overcome this issue, in this paper, we present an AutoML-based model to classify the raw UV-Vis spectroscopy data. AutoML \cite{hutter2019automated} is a machine learning technique used to discover the best-performing pipeline of data transforms, models, and model configurations for a dataset.
It is designed to reduce the demand for data scientists and to enable domain experts to automatically build ML applications without much prerequisite statistical and ML knowledge. As it's pipeline architecture comprises of all essential components to build a model, it holds several advantages over the conventional classification models. AutoML provides automated hyperparameter optimization which reduces human effort and improves the reproducibility and fairness of scientific studies. 
The meta-learning component exploits the ability of computer systems to store extensive amounts
of prior learning experiences which can be leveraged to learn new tasks. AutoML employs neural architecture search (NAS) that creates a search space, search strategy, and performance estimation strategy to find architectures that achieve high predictive performance on unseen data. We compared the performance of four AutoML models, Auto-SKlearn \cite{feurer2019auto}, TPOT \cite{olson2016evaluation}, Hyperopt-SKlearn \cite{komer2014hyperopt}, and AutoKeras \cite{jin2019auto}, to determine the most suitable model for SCT/SCD screening based on spectroscopy data. The proposed AutoML model can complement the absorbance-based test to rapidly screen for SCT/SCD and enable the deployment of the test in resource-limited settings.

The paper is organized as follows. Firstly, we describe the dataset and the data preprocessing method adopted for extraction of relevant features in section \ref{sec:method}. The various models used for the study and the parameter tuning for each model are described in section \ref{ssec:model}. The performance of the various models in terms of various performance metrics is described in section \ref{sec:result}.

\section{Methods}
\label{sec:method}

\subsection{Data Collection}
\label{ssec:subhead}
In the spectrocopy test proposed by Rajesh et al. \cite{rajesh}, the blood samples are mixed in a special phosphate buffer containing a lysing agent and a reducing agent and incubated for 15 minutes. The UV-Vis absorbance spectra of the solution is then captured for each sample. The total absorbance spectroscopy test dataset includes 199 normal samples, 16 sickle cell disease samples, and 156 sickle cell trait samples. The spectroscopy-based test was performed on blood samples collected from the tribal regions of Madhya Pradesh, India, where sickle cell anemia is endemic.  The spectroscopy test was performed on the samples using the low-cost, portable Prizm Plus spectrometer (Testright systems, India). All samples were tested using HPLC to obtain the ground truth.
  
%
%
%

\begin{figure}[h!]
\centering
\includegraphics[width=3 in]{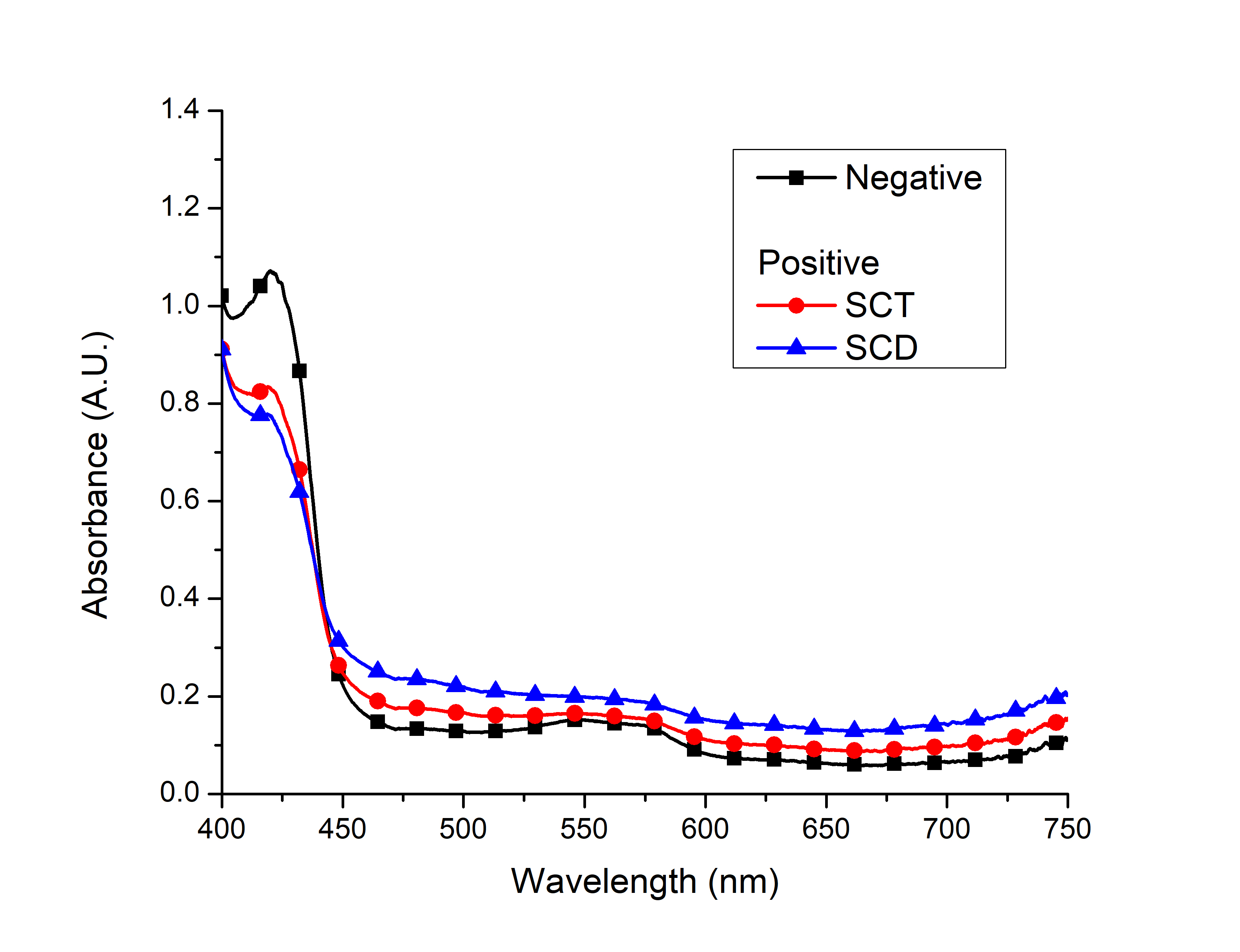}
\caption{Absorbance spectra of negative and positive samples. In this study, both SCT and SCD samples are considered as positive samples.}
\label{5}
\end{figure}

\subsection{Data preparation}
\label{ssec:model}

The absorbance spectra of the positive and negative samples are shown in Fig. \ref{5}. For each sample, the absorbance data was captured in the wavelength range from 395 nm to 750 nm, with a resolution of 0.165 nm. Here, we refer to a wavelength and its corresponding absorbance value as a feature. Hence, the data corresponding to each sample consisted of 2183 features. As this would significantly increase the complexity of the model, we sought to reduce the number of features of each sample data.  From careful observation of the absorbance spectra, we decided to consider only the absorbance spectra in the range of 400 nm to 600 nm for our analysis. Outside this range, the absorbance values are mostly constant, providing no useful insight. Then, the data was averaged over each wavelength to form a dataset with integral wavelengths (1 nm resolution), as the variation in the absorbance values for sub-nm wavelengths was observed to be minimal. Further, the data was averaged over 5 nm-intervals in order to create the final sample data. Thus, in the final dataset used to train the models, each sample data had 41 features.

\subsection{Data split}
\label{ssec:split}

The data was split between training and testing sets in the 70:30 ratio. The size of the datasets are summarized below in Table 1.

\begin{table}[h]
\centering
\resizebox{0.28\textwidth}{!}{%
\begin{tabular}{l c c c c }
\hline 
 Sample & Train & Test  & Total \\ 
\hline
Normal & 134 & 65  & 199 \\
SCT/SCD & 125 & 47 & 172\\
\hline
\end{tabular}}
\caption{Data distribution across various sets}
\end{table}

\subsection{AutoML Models}
\label{ssec:model}

We evaluated the classification performance of four different AutoML models to find the most suitable model for the screening of sickle hemoglobin positive samples.
 
\subsection{Auto-Sklearn}
\label{ssec:autosklearn}
Auto-sklearn is a system based on scikit-learn and uses 15 classifiers, 14 feature preprocessing methods, and 4 data preprocessing methods, giving rise to a structured hypothesis space with 110 hyperparameters. It improves on existing AutoML methods by automatically taking into account past performance on similar datasets and by constructing ensembles from the models evaluated during the optimization. After multiple trials, for optimal results, the execution time was set to 15 minutes and the metric was chosen as precision\_micro.

\subsection{TPOT}
\label{ssec:tpot}
Tree-based pipeline optimization tool (TPOT) is a model that automatically designs and optimizes machine learning pipelines. TPOT uses a version of genetic programming to automatically design and optimize a series of data transformations and machine
learning models that attempt to maximize the classification accuracy for a given supervised learning data set. Optimal results were obtained when we set the number of generations to 5, population size to 100, and scoring to be precision\_micro.
 
\subsection{Hyperopt-Sklearn}
\label{ssec:hyperopt}
Hyperopt-sklearn provides a unified interface to a large subset of
the machine learning algorithms available in scikit-learn. With the help of its optimization functions, Hyperopt is able to search over all possible configurations of scikit-learn components, including preprocessing, classification, and regression modules to surpass human experts in algorithm configuration. 
In this study, optimal results were obtained when SVC was chosen as the classifier with max\_evals set to 70 and trail\_timeout to 60.

\subsection{Auto-Keras}
\label{ssec:autokeras}

Auto-Keras uses Bayesian optimization to guide the search by designing a neural network kernel and an algorithm for optimizing acquisition function in tree-structured space, wrapped on to the Keras package. After multiple experiments, we set the max trials to 20 and epochs to 45. The model was set to iterate over 20 different Keras models trained for 20 epochs each.

\section{Result}
\label{sec:result}

\subsection{Overall Performance}
\label{ssec:overall}

The parameters of each model were tuned over several iterations to achieve the best model performance.
The overall performance of the models is reported in Table 2. The models were evaluated on the basis of sensitivity, specificity, accuracy, positive predictive value (PPV), and negative predictive value (NPV).  Here, accuracy is defined as $\frac{TP+TN}{TP+FN+TN+FP}$, where TP is true positive, FN is false negative, TN is true negative, and FP is false positive. Sensitivity is defined as $\frac{TP}{TP+FN}$, specificity is defined as $\frac{TN}{TN+FP}$, PPV is defined as $\frac{TP}{TP+FP}$, and NPV is defined as $\frac{TN}{TN+FN}$.

\begin{table}[h]
\resizebox{0.49\textwidth}{!}{%
\begin{tabular}{l c c c c}
\hline 
 Metric & Auto-sklearn & TPOT & Hyperopt-Sklearn & Auto-Keras \\ 
\hline
TP & 45 & 47 & 46 & 47\\
TN & 63 & 61 & 61 & 59\\
FP & 2  & 4  & 4  & 6 \\
FN & 2  & 0  & 1  & 0 \\
Sensitivity & 95.74 & 100 & 97.87 & 100\\
Specificity & 96.92 & 93.84 & 93.84 & 90.76 \\
Accuracy & 96.43 & 96.43 & 95.53 & 94.64\\
PPV & 95.74 & 92.15 & 92.00 & 88.67 \\
NPV & 96.92 & 100 & 98.38 & 100 \\
\hline
\end{tabular}}
\caption{Performance metrics of the various models on the test dataset}
\end{table}


In addition to the performance metrics, another parameter that should be taken into consideration while evaluating the performance of the diagnostic test is the number of false negatives. A false negative test, in the case of sickle cell screening, can lead to the patient not receiving genetic counselling and lead to further generational propagation of sickle cell disease. Hence, an ideal diagnostic test should have zero false negatives. In our study, both TPOT and AutoKeras showed zero false negatives. However, the TPOT outperforms all the other models in terms of all the performance metrics as shown in Table 2. Therefore, we conclude that TPOT is the best performing model that can screen for SCT/SCD with 100\% sensitivity. In future, we intend to develop a machine learning based test that can distinguish between SCD and SCT. In this study, we could only obtain 16 SCD patient blood samples. On significantly increasing the number of SCD samples used for training the AutoML model, it would be possible to classify SCT and SCD samples accurately. We also intend to train the classifier on data obtained from multiple spectrometer instruments to account for the effect of instrument variations on the classification performance.

\section{Conclusion}
\label{sec:conclusion}

In summary, it is demonstrated that the proposed AutoML-based technique applied on the UV-Vis spectroscopy data can screen for sickle cell trait and disease with high accuracy, sensitivity, and specificity. After training, the proposed model can be deployed on a mobile phone app to enable local health workers to interpret the data from the low-cost spectroscopy-based test in resource-limited clinical settings. In future work, we intend to create a ML model that can differentiate between trait and disease conditions, which can potentially replace the conventional confirmatory tests, such as HPLC and electrophoresis.

\bibliographystyle{IEEEbib}
\bibliography{Template_ISBI_latex}

\begin{thebibliography}{10}

\bibitem{kato2018sickle}
Gregory~J Kato, Fr{\'e}d{\'e}ric~B Piel, Clarice~D Reid, Marilyn~H Gaston,
  Kwaku Ohene-Frempong, Lakshmanan Krishnamurti, Wally~R Smith, Julie~A
  Panepinto, David~J Weatherall, Fernando~F Costa, et~al.,
\newblock ``Sickle cell disease,''
\newblock {\em Nature Reviews Disease Primers}, vol. 4, no. 1, pp. 1--22, 2018.

\bibitem{tubman2015sickle}
Ven{\'e}e~N Tubman and Joshua~J Field,
\newblock ``Sickle solubility test to screen for sickle cell trait: what's the
  harm?,''
\newblock {\em Hematology 2014, the American Society of Hematology Education
  Program Book}, vol. 2015, no. 1, pp. 433--435, 2015.

\bibitem{piety2017paper}
Nathaniel~Z Piety, Alex George, Sonia Serrano, Maria~R Lanzi, Palka~R Patel,
  Maria~P Noli, Silvina Kahan, Damian Nirenberg, Jo{\~a}o~F Camanda, Gladstone
  Airewele, et~al.,
\newblock ``A paper-based test for screening newborns for sickle cell
  disease,''
\newblock {\em Scientific reports}, vol. 7, no. 1, pp. 1--8, 2017.

\bibitem{yang2013simple}
Xiaoxi Yang, Julie Kanter, Nathaniel~Z Piety, Melody~S Benton, Seth~M Vignes,
  and Sergey~S Shevkoplyas,
\newblock ``A simple, rapid, low-cost diagnostic test for sickle cell
  disease,''
\newblock {\em Lab on a Chip}, vol. 13, no. 8, pp. 1464--1467, 2013.

\bibitem{nnodu2019hemotypesc}
Obiageli Nnodu, Hezekiah Isa, Maxwell Nwegbu, Chinatu Ohiaeri, Samuel Adegoke,
  Reuben Chianumba, Ngozi Ugwu, Biobele Brown, John Olaniyi, Emmanuel Okocha,
  et~al.,
\newblock ``Hemotypesc, a low-cost point-of-care testing device for sickle cell
  disease: promises and challenges,''
\newblock {\em Blood Cells, Molecules, and Diseases}, vol. 78, pp. 22--28,
  2019.

\bibitem{rajesh}
Rajesh Srinivasan, Eugene Christo V.~R., Prateek Katare, Aravind Venukumar,
  Nisanth K~M Nambison, and Sai~Siva Gorthi,
\newblock ``Methods for identifying haemoglobin {S} or {C} in a biological
  sample and kits thereof,'' India Patent 202141009024, March 2021.

\bibitem{hutter2019automated}
Frank Hutter, Lars Kotthoff, and Joaquin Vanschoren,
\newblock {\em Automated machine learning: methods, systems, challenges},
\newblock Springer Nature, 2019.

\bibitem{feurer2019auto}
Matthias Feurer, Aaron Klein, Katharina Eggensperger, Jost~Tobias Springenberg,
  Manuel Blum, and Frank Hutter,
\newblock ``Auto-sklearn: efficient and robust automated machine learning,''
\newblock in {\em Automated Machine Learning}, pp. 113--134. Springer, Cham,
  2019.

\bibitem{olson2016evaluation}
Randal~S Olson, Nathan Bartley, Ryan~J Urbanowicz, and Jason~H Moore,
\newblock ``Evaluation of a tree-based pipeline optimization tool for
  automating data science,''
\newblock in {\em Proceedings of the genetic and evolutionary computation
  conference 2016}, 2016, pp. 485--492.

\bibitem{komer2014hyperopt}
Brent Komer, James Bergstra, and Chris Eliasmith,
\newblock ``Hyperopt-sklearn: automatic hyperparameter configuration for
  scikit-learn,''
\newblock in {\em ICML workshop on AutoML}. Citeseer, 2014, vol.~9, p.~50.

\bibitem{jin2019auto}
Haifeng Jin, Qingquan Song, and Xia Hu,
\newblock ``Auto-keras: An efficient neural architecture search system,''
\newblock in {\em Proceedings of the 25th ACM SIGKDD International Conference
  on Knowledge Discovery \& Data Mining}, 2019, pp. 1946--1956.

\end{thebibliography}

\end{document}